\documentclass[aps,prl,amsmath,amssymb,floatfix,twocolumn,amsmath,superscriptaddress,twocolumn,nofootinbib,tighten,letterpaper]{revtex4-1}
\usepackage{multirow}
\usepackage{bbold}

\usepackage{color}
\usepackage{mathrsfs}
\usepackage{hyperref}
\usepackage[normalem]{ulem}
\usepackage{bm}

\usepackage[caption=false]{subfig} 

\usepackage{amsfonts, relsize, color}
\usepackage{graphics}
\usepackage{graphicx}
\usepackage{subfloat}

\usepackage{hyperref}
\usepackage{color}
\usepackage{comment}

\renewcommand\vec[1]{\ensuremath\boldsymbol{#1}} 

\begin{document}
\title{Competing orders and cascade of degeneracy lifting in doped Bernal bilayer graphene}

\author{Andr$\acute{\mbox{a}}$s L. Szab$\acute{\mbox{o}}$}
\affiliation{Max-Planck-Institut f\"{u}r Physik komplexer Systeme, N\"{o}thnitzer Str. 38, 01187 Dresden, Germany}

\author{Bitan Roy}\thanks{Corresponding author: bitan.roy@lehigh.edu}
\affiliation{Department of Physics, Lehigh University, Bethlehem, Pennsylvania, 18015, USA}

\date{\today}
\begin{abstract}
Motivated by recent experiments [H. Zhou, \emph{et al.}, Science {\bf 375}, 774 (2022) and S. C. de la Barrera, \emph{et al.}, arXiv:2110.13907], here we propose a general mechanism for valley and/or spin degeneracy lifting of the electronic bands in doped Bernal bilayer graphene, subject to electric displacement ($D$) fields. A $D$-field induced layer polarization (LP), when accompanied by Hubbard repulsion driven layer antiferromagnet (LAF) and next-nearest-neighbor repulsion driven quantum anomalous Hall (QAH) orders, lifts the four-fold degeneracy of electronic bands, yielding a quarter metal for small doping, as also observed in ABC trilayer graphene. With the disappearance of the QAH order, electronic bands recover two-fold valley degeneracy, thereby forming a conventional or compensated (with majority and minority carriers) half-metal at moderate doping, depending on the relative strength of LP and LAF. At even higher doping and for weak $D$-field only LAF survives and the Fermi surface recovers four-fold degeneracy. We also show that a pure repulsive electronic interaction mediated triplet $f$-wave pairing emerges from a parent correlated nematic liquid or compensated half-metal when an in-plane magnetic field is applied to the system.  
\end{abstract}

\maketitle

\begin{figure}[t!]
\includegraphics[width=0.90\linewidth]{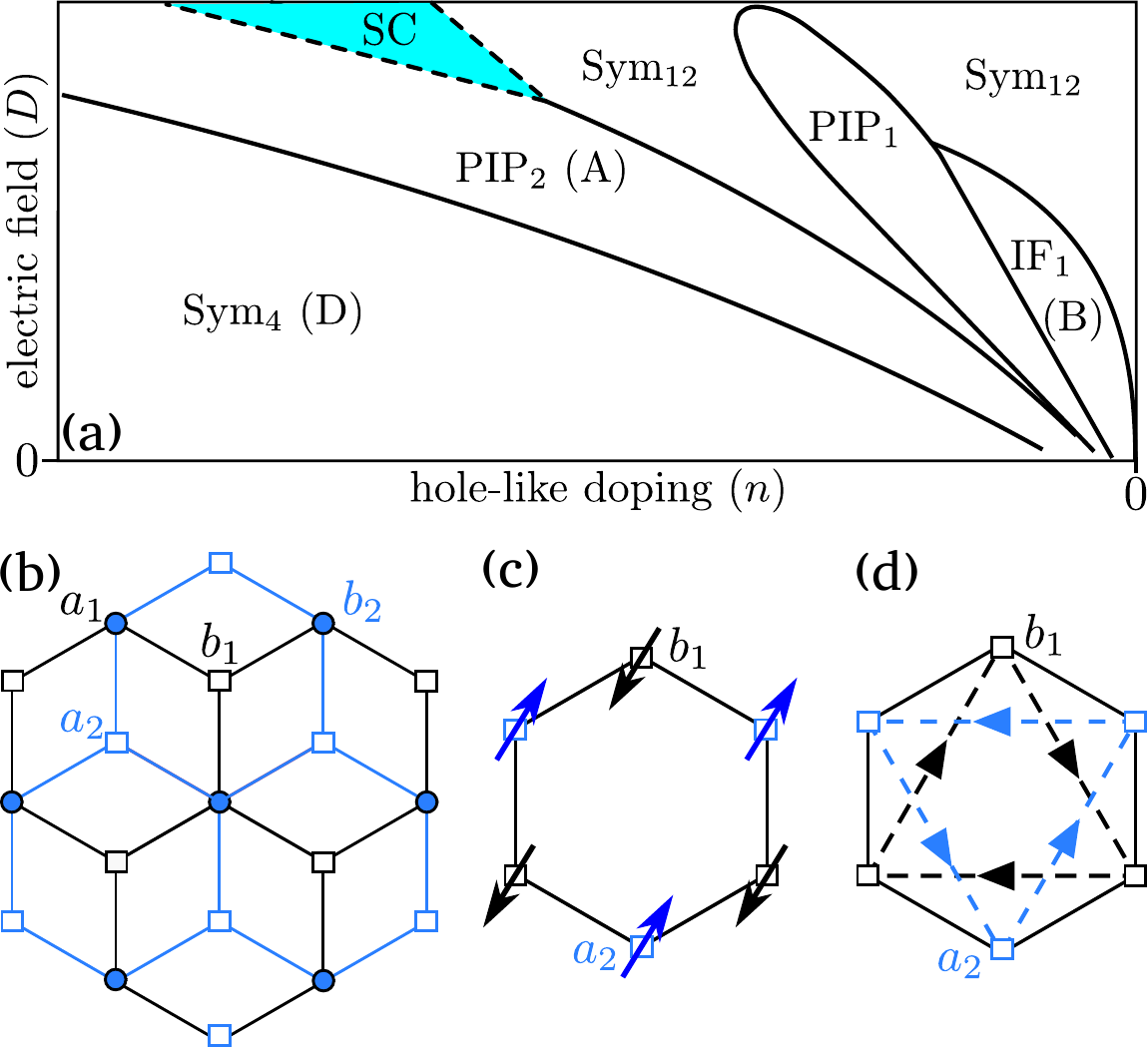}
\caption{(a) Schematic phase diagram of BBLG~\cite{BLGExpRec:1, BLGExpRec:2}. For symbols of various phases see text. The superconductor (SC) is found only in the presence of an in-plane magnetic field. (b) Top view of BBLG. The subscript $i=1,2$ is the layer index of the sites. Each $a_1$ and $b_2$ sites overlap. Eigenstates of the high energy split-off bands reside dominantly on these two dimer sites. Sites $a_2$ and $b_1$ form a honeycomb lattice and participate in the low-energy description of BBLG. (c) LAF with electronic spins on the $a_2$ and $b_1$ sites pointing in the opposite directions. (d) QAH order with intrasublattice circulating currents in opposite directions on two layers.  
}~\label{fig:orders}
\end{figure}

\emph{Introduction}.~Altogether an isolated layer of carbon-based honeycomb membrane and its stacked cousins constitute a diverse zoo of gapless chiral quasiparticles, displaying a variety of nodal electronic band dispersion~\cite{graphene:RMP}. While monolayer graphene hosts quasirelativistic Dirac fermions~\cite{MLG:band1, MLG:band2}, gapless electronic bands acquire quadratic and cubic dispersions in Bernal bilayer graphene (BBLG)~\cite{BLG:band} and ABC or rhombohedral trilayer graphene (RTLG)~\cite{RTLG:band}, respectively. Furthermore, a relative twist between two layers of graphene substantially reduces the Fermi velocity of Dirac fermions near the so-called magic angle~\cite{TBLGband:1, TBLGband:2, TBLGband:3, TBLGband:4}, where a cascade of insulating and superconducting phases has been observed by tuning the carrier density~\cite{TBLGSC:1, TBLGSC:2, TBLGSC:3, TBLGSC:4}. A similar phenomenon has also been reported in RTLG~\cite{young:RTLG1, young:RTLG2}, and very recently in BBLG~\cite{BLGExpRec:1, BLGExpRec:2} [Fig.~\ref{fig:orders}(a)]. Although interaction driven competing orders in RTLG have been investigated theoretically~\cite{RTLGrecent:0, RTLGrecent:1, RTLGrecent:2, RTLGrecent:3, RTLGrecent:4, RTLGrecent:5, RTLGrecent:6}, a similar expedition exploring the landscape of doped BBLG, subject to an external electric displacement ($D$) field, is yet to be initiated. This Letter is geared toward unveiling the nature of underlying competing orders, causing systematic degeneracy lifting in BBLG and the appearance of proximal superconductivity.

\emph{Summary}.~First we review the key experimental observations and summarize our main results. In BBLG subject to $D$-field, quantum oscillations show a single peak at frequency (in units of $n \phi_0$) $\nu=1$ at low carrier density $n$. Here $\phi_0$ is the flux quanta. The corresponding broken symmetry phase (IF$_1$~\cite{BLGExpRec:1} or B (C) on the hole (electron) dope side~\cite{BLGExpRec:2}), thus lacks spin and valley degeneracy. The adjacent PIP$_1$ phase at slightly higher doping shows a quantum oscillation peak at $\nu \lesssim 1$.

\begin{figure}[t!]
\includegraphics[width=0.98\linewidth]{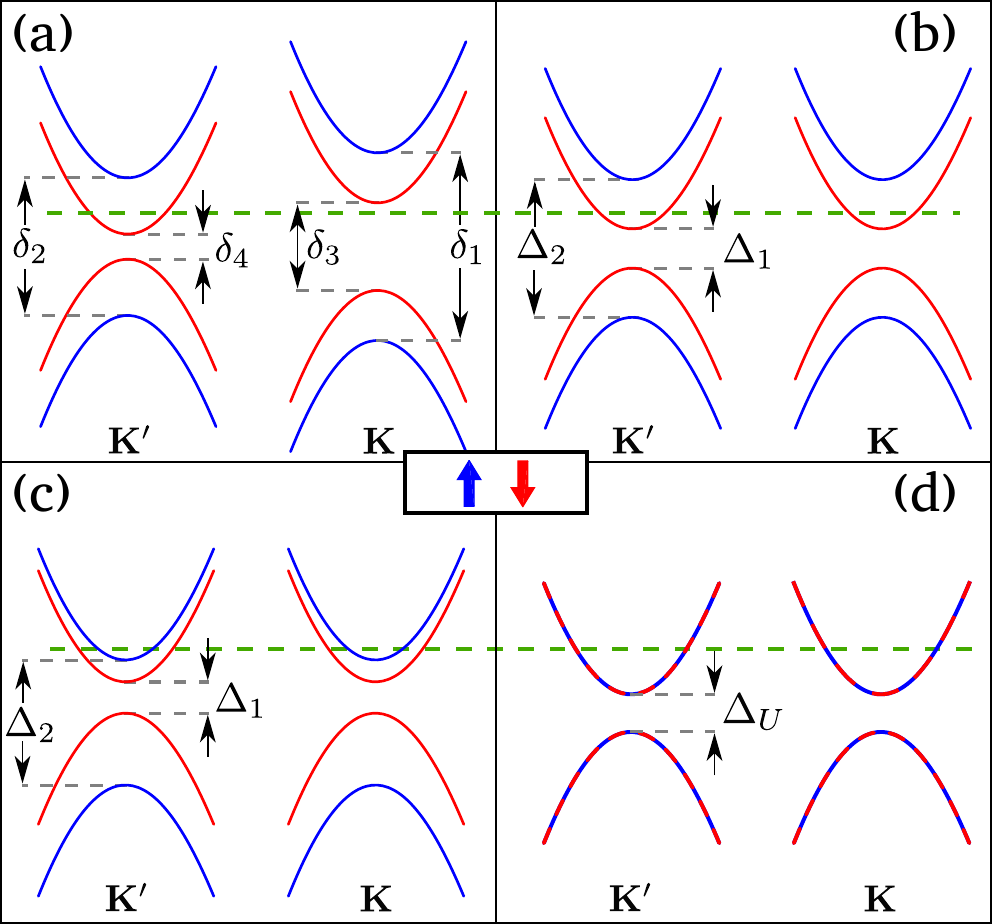}
\caption{(a) Quarter-metal for low chemical potential (green line) in the presence of induced LP, interaction-driven LAF and QAH orders, lifting valley and spin degeneracies. For fixed chemical potential but in the absence of QAH order, valley degeneracy is restored, yielding (b) regular and (c) compensated (with majority and minority carriers) half-metals, respectively for large and small LAF order. (d) In the absence of LP, spin degeneracy is recovered, yielding a metal. Here ${\bf K}$ and ${\bf K}^\prime$ are two valleys. For other symbols see text.    
}~\label{fig:metal}
\end{figure}

We show that the four-fold degeneracy of electronic bands (stemming from the valley and spin degrees of freedom) is completely lifted when a $D$-field induced layer polarization (LP) is accompanied by Hubbard repulsion ($U$) driven layer antiferromagnet (LAF) and intralayer next-nearest-neighbor repulsion ($V_2$) mediated quantum anomalous Hall (QAH) orders. The LAF leads to staggered pattern of electronic spin between the low-energy sites of BBLG residing on complementary layers, while the QAH order supports circulating currents among the intralayer low-energy sites, with opposite orientations on two layers~\cite{nandkishore:BLG, lemonik:BLG, macdonald:BLG, vafek:BLG, roy:BLG}. See Fig.~\ref{fig:orders}. As the resulting Fermi surface contains only one out of four degrees of freedom, contributing to quantum oscillations, this phase is named \emph{quarter-metal} [Fig.~\ref{fig:metal}(a)]. An \emph{identical} mechanism can be responsible for the observed quarter-metal in RTLG~\cite{young:RTLG1, RTLGrecent:5}. The adjacent PIP$_1$ phase can be realized by injecting carriers to the quarter-metal, such that a small Fermi surface develops for another spin or valley degrees of freedom.

At moderate hole doping, quantum oscillations show two peaks at $\nu=\nu_{_1}$ and $\nu_{_2}$, with $\nu_{_1}+\nu_{_2}=1/2$, as in Ref.~\cite{BLGExpRec:1} (phase PIP$_2$), possibly also in Ref.~\cite{BLGExpRec:2} (phase A). Thus, such a phase features two-fold degenerate Fermi surfaces, however with majority and minority carriers.

With the disappearance of QAH order, residual LP and LAF yield a valley-degenerate half-metal with broken spin degeneracy [Fig.~\ref{fig:metal}(b)] that shows a single oscillation peak at $\nu=1/2$. However, depending on the relative strength of LP and LAF, the system can also form a compensated half-metal with majority (large Fermi surface) and minority (small Fermi surface) carriers for opposite spin projections [Fig.~\ref{fig:metal}(c)], producing two oscillation peaks at frequencies $\nu_1$ and $\nu_2$, such that $\nu_1+\nu_2=1/2$.

At even larger doping and weak $D$-field, one oscillation peak appears at $\nu=1/4$. This phase (Sym$_4$~\cite{BLGExpRec:1} or D~\cite{BLGExpRec:2}), preserving the four-fold degeneracy, can result from a pure LAF order [Fig.~\ref{fig:metal}(d)].

Another phase (Sym$_{12}$), displaying oscillation peak at $\nu=1/12$ appears at small and moderate doping, separated by PIP$_1$. A three-fold rotational symmetry breaking nematic order, preserving four-fold degeneracy stands as its candidate, which at moderate (low) doping is possibly stabilized by electronic interaction ($D$-field~\cite{graphene:RMP}).

\begin{figure}[t!]
\includegraphics[width=0.97\linewidth]{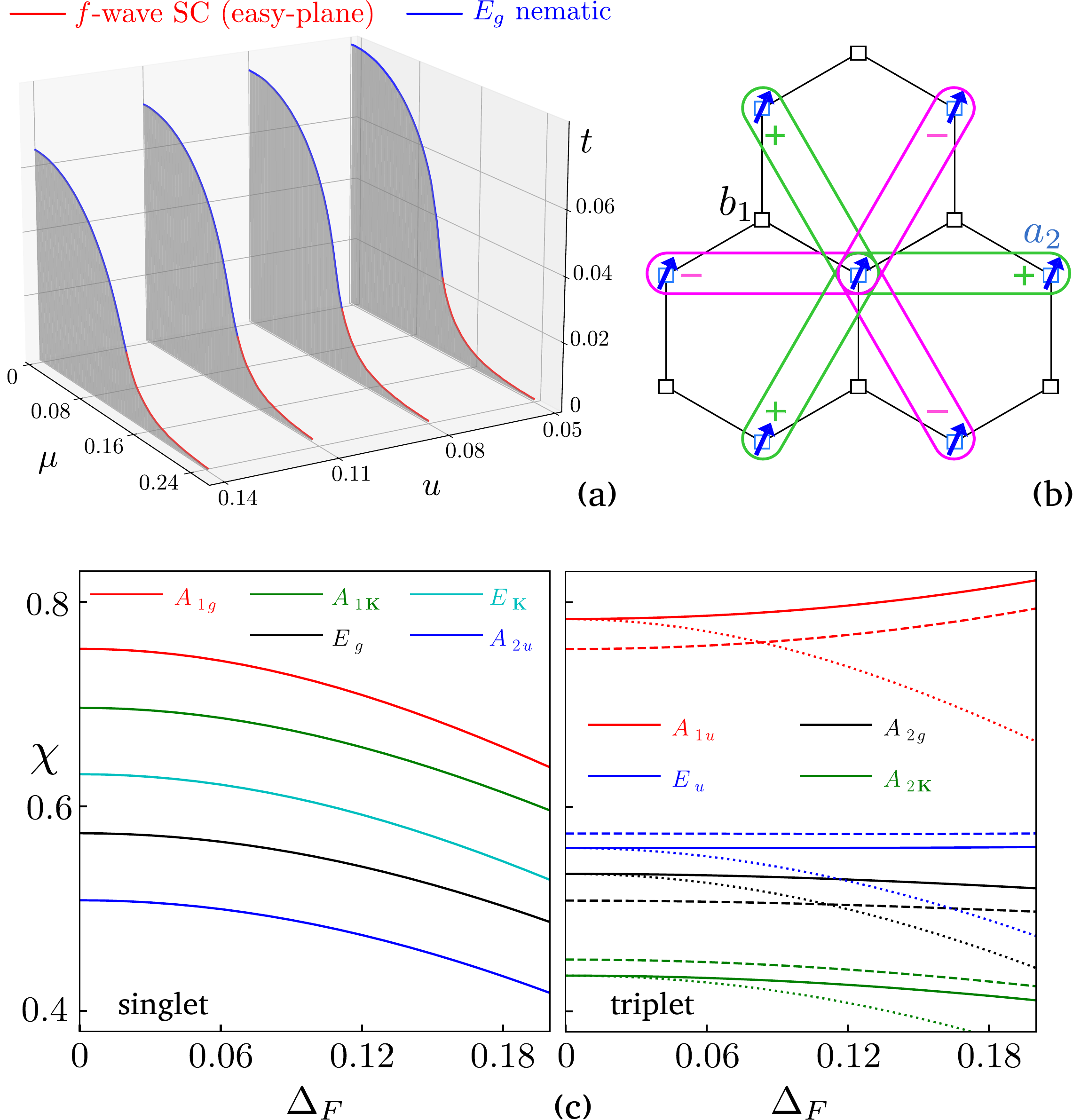}
\caption{(a) Various cuts of the phase diagram for fixed bare interaction $\lambda{_{E_g}}=0.03$ and Zeeman coupling $h=0.3$, showing a confluence of $E_g$ nematicity and easy-plane $f$-wave pairing. The shaded (white) region represents ordered (disordered) phase. (b) Intralayer $f$-wave Cooper pairs. (c) Pairing susceptibility $\chi$ [Eq.~(\ref{eq:susceptibility})] in a compensated half-metal for $\mu=0.5$, $t=0.2$, $u=0.25$ and $\Delta_{A}=0.125$. For each triplet channel solid line corresponds to the component perpendicular to LAF and $B_\parallel$-field, which for the $f$-wave pairing possesses the largest susceptibility and is thus a candidate for the observed pairing. The dashed (dotted) lines represent components parallel to LAF ($B_\parallel$-field). All parameters are dimensionless. 
}~\label{fig:superconductivity}
\end{figure}

Finally, with the application of an in-plane magnetic field ($B_\parallel$), a superconductor appears in between PIP$_2$ and the correlated nematic liquid (Sym$_{12}$)~\cite{BLGExpRec:1}. This paired state is possibly spin-triplet in nature, as it exceeds the Pauli-limiting $B_\parallel$-field. We show that repulsive electronic interaction in the nematic channel is conducive for the nucleation of a spin-triplet $f$-wave pairing, with its spin component being locked in the easy-plane, perpendicular to the $B_\parallel$-field. The proximal compensated half-metal (PIP$_2$) also favors such $f$-wave pairing. See Fig.~\ref{fig:superconductivity}.

\emph{Model}.~We arrive at these conclusions from the low-energy model for BBLG. The unit cell is composed of four sites. Each layer contributes to two of them [Fig.~\ref{fig:orders}]. With no interlayer hopping, each layer hosts massless Dirac fermions. The direct interlayer hopping between the dimer $b_1$ and $a_2$ sites ($t_\perp \approx 200$meV), where the subscript corresponds to the layer index, couples Dirac fermions as static non-Abelian magnetic field~\cite{BBLG:nonabelian}. It pushes two out of four bands to high-energies (split-off bands). The remaining two bands display bi-quadratic touchings at two inequivalent corners of the Billouin zone, giving rise to the valley or isospin degrees of freedom. Wavefunctions for the low-energy bands reside predominantly on the $a_1$ and $b_2$ sites. Accounting for the layer, valley and spin degrees of freedom, we arrive at the low-energy model for doped BBLG subject to a $D$-field 
\begin{equation}~\label{eq:hamilBBLG}
H_0= \frac{1}{{2 m_\star}}\left[ d_1 (\vec{k}) \Gamma_{3001} - d_2(\vec{k}) \Gamma_{3032} \right] + u \Gamma_{3003} -\mu \Gamma_{3000}.
\end{equation} 
Here $d_1 (\vec{k})=k^2_x-k^2_y$, $d_2 (\vec{k})=2 k_x k_y$, momentum $\vec{k}$ is measured from the respective valleys, $m_\star \approx 0.028 m_e$ is the effective mass of quasiparticle excitations, where $m_e$ is the mass of free electrons, $u=-D d_0$ is the external $D$-field induced LP with $d_0$ as interlayer separation, and $\mu$ is the chemical potential measured from the charge neutrality point. The sixteen-dimensional Hermitian matrices are $\Gamma_{\mu \nu \rho \lambda}=\eta_\mu \sigma_\nu \tau_\rho \beta_\lambda$. Four sets of Pauli matrices $\{\eta_\mu \}$, $\{ \sigma_\nu \}$, $\{ \tau_\rho \}$ and $\{ \beta_\lambda \}$ respectively operate on the particle-hole, spin, valley and layer indices, with $\mu, \nu, \rho, \lambda=0, \dots, 3$. We introduce Nambu doubling to capture superconductivity~\cite{supplementary}. Here we neglect the intralayer nearest-neighbor (NN) repulsion, as all the NN sites of any low-energy site in each layer ($b_1$ and $a_2$) are dimer sites ($a_1$ and $b_2$), which do not participate in the low-energy theory. The interlayer NN repulsion can be neglected as the $D$-field already breaks the layer inversion symmetry of electronic density, yielding a LP state.

We neglect the particle-hole asymmetry as it does not play any role in the pattern of symmetry breaking. We also neglect the trigonal warping, which splits quadratic band touching points into Dirac points. As the density of states for quadratic (linear) band dispersion is constant (vanishes linearly) with energy in two-dimensions, the effects of short-range Coulomb interactions, responsible for spontaneous symmetry breaking, are primarily driven by the component quadratic in $\vec{k}$. Our proposed candidates for half- and quarter-metals, and the paired state, however, remain unaffected by trigonal warping~\cite{supplementary}.

\emph{Quarter-metal}.~On site Hubbard repulsion is expected to be the dominant component of finite range Coulomb interaction even in BBLG~\cite{katsnelson:hubbard}. It favors LAF, which when simultaneously present with LP, lifts the spin degeneracy~\cite{RTLGrecent:5}. Although next-nearest-neighbor repulsion favors quantum spin Hall order, once the spin degeneracy is lifted, the system behaves as a spinless one, and favors QAH order~\cite{szaboroy:selection}. In the presence of LP, LAF and QAH orders, the effective single-particle Hamiltonian is
\allowdisplaybreaks[4]
\begin{equation}
H_{\rm QM}=H_0 + \Delta_U \Gamma_{0303} + \Delta_V \Gamma_{0033}.
\end{equation}
Here $\Delta_U \sim U$ and $\Delta_V \sim V_2$ are the amplitudes of the LAF and QAH orders, respectively. The corresponding energy spectra are $\pm \sqrt{\xi^2_{\vec k} + \delta^2_i}-\mu$ for $i=1,\dots, 4$, where  
\begin{equation}
\delta_1-u=-\delta_4+u=\Delta_1+\Delta_2, \: 
\delta_2-u=-\delta_3+u=\Delta_1-\Delta_2, \nonumber 
\end{equation}
and $\xi_{\vec{k}}=|\vec{k}|^2/(2 m_\star)$. As the four-fold spin and valley degeneracy is completely lifted, the system behaves like a quarter-metal, when $|\delta_4| < \mu < |\delta_3|$, supporting a nondegenerate Fermi surface [Fig.~\ref{fig:metal}(a)], as observed in experiments~\cite{BLGExpRec:1, BLGExpRec:2}. The key mechanism behind the formation of a quarter-metal is the simultaneous presence of \emph{three} \emph{masses} (LP, LAF and QAH), each of which \emph{anticommutes} with the $\vec{k}$-dependent part of the Hamiltonian $H_0$, but mutually \emph{commute} with each other. When $|\delta_3| < \mu < |\delta_2|$, the system supports large and small Fermi surfaces, featuring oscillation peak at $\nu \lesssim 1$ as in the PIP$_1$ phase~\cite{BLGExpRec:1}. Next we generalize this mechanism for a (compensated) half-metal.

\emph{Half-metal}.~Chemical potential yields an infrared cutoff for the renormalization group (RG) flow of four-fermion interactions. Thus, although at the bare level $U>V_2$~\cite{interactionstrength:comment}, at low $\mu$ their renormalized strengths diverge in a similar fashion, as all quartic interactions have the same scaling dimension and receive similar quantum corrections, which wash out their difference at the microscopic or bare level after long RG time once many fermionic modes are integrated out. As a result the LAF and QAH orders show a prominent competition in the infrared regime, giving rise to a quarter-metal. But, at intermediate $\mu$ when only a few fermionic modes are integrated out, Hubbard-U dominates over $V_2$ as $U>V_2$ at the microscopic level and quantum corrections are insufficient to smear out such difference, resulting in the disappearance of the QAH order. Also, at the bare level Hubbard-U is attractive in the LAF and ferromagnet channels, as shown from the generalized Hamman decomposition~\cite{szaboroy:selection}. But, as the system flows into the low $\mu$ regime, the ferromagnetic component disappears under coarse grain, and only LAF order survives, as for the quarter-metal. By contrast, at intermediate $\mu$, LAF can in principle be accompanied by at least a small ferromagnet ($\Delta_F$). Then the effective single-particle Hamiltonian reads
\begin{equation}
H_{\rm CHF}=H_0 + \Delta_U \Gamma_{0103} + \Delta_F \Gamma_{0300}.
\end{equation}          
The relative spin orientations of the ferromagnet and LAF are chosen in such a way that their corresponding matrix operators mutually anticommute, yielding a configuration of maximal condensation energy gain. The energy spectra of $H_{\rm CHF}$ are $\pm E_{\tau, \sigma}-\mu$, where for $\tau, \sigma=\pm$ 
\begin{equation}
E_{\tau, \sigma}=\left[ z^2 + \Delta^2_U + \Delta^2_F + 2 \sigma \sqrt{ u^2 \Delta^2_U + \Delta^2_F z^2} \right]^{1/2}
\end{equation}     
and $z^2=\xi^2_{\vec{k}} + u^2$. The independence of $E_{\tau, \sigma}$ on $\tau$ reflects valley degeneracy of each band. When $\Delta_1<\mu<\Delta_2$, the system describes a half-metal [Fig.~\ref{fig:metal}(b)], with one type of carriers, featuring quantum oscillations at $\nu=1/2$, where $\Delta_1=u-\sqrt{\Delta^2_U+\Delta^2_F}$ and $\Delta_2=u+\sqrt{\Delta^2_U+\Delta^2_F}$. However, for sufficiently small $\sqrt{\Delta^2_U+\Delta^2_F}$, it is conceivable that $\mu> \Delta_{1,2}$ for the same fixed $\mu$ [Fig.~\ref{fig:metal}(c)]. The system then describes a compensated half-metal, as depletion of majority carriers for one spin projection is compensated by the equal gain of minority carriers for opposite spin projection. This phase shows two quantum oscillation peaks at $\nu_{_1}$ and $\nu_{_2}$, with $\nu_{_1}+\nu_{_2}=1/2$. Although we can arrive at the same conclusions for $\Delta_F=0$, shortly we justify the presence of ferromagnet.

\emph{Metal}.~A new phase (Sym$_4$~\cite{BLGExpRec:1} or D~\cite{BLGExpRec:2}) sets in via a phase transition from the compensated half-metal, triggered by decreasing $D$-field or LP. It shows quantum oscillation peak at $\nu=1/4$, stemming from four-fold degenerate Fermi surface. The $D$-field acts as an infrared cutoff for the RG flow of four-fermion interactions. Thus, as the $D$-field decreases, Hubbard-U undergoes more coarse grain, leading to the disappearance of the ferromagnet. Since the Hubbard-U dominates in BBLG, we believe this phase possesses pure LAF order [Fig.~\ref{fig:metal}(d)].

The identification of distinct orderings leading to degeneracy lifting in BBLG qualitatively agrees with the evolution of the phase boundaries among them in the presence of $B_\parallel$-field~\cite{BLGExpRec:2}. Its dominant effect is the Zeeman coupling with electronic spin~\cite{TBLGband:inplanefield}, naturally boosting the ferromagnet component of the compensated half-metal, but detrimental to LAF order. Consequently, with increasing $B_\parallel$-field the compensated half-metal is found over a larger doping range~\cite{BLGExpRec:2}.

\emph{Superconductivity}.~With the application of a $B_\parallel$-field, a wedge shaped superconducting regime appears in between Sym$_{12}$ and PIP$_2$~\cite{BLGExpRec:1}. To shed light on such paired state we perform a leading-order RG analysis by considering a repulsive ($g_{_{E_g}}>0$) four-fermion interaction~\cite{interactionstrength:comment}    
\begin{equation}
g_{_{E_g}} \left[ \left( \Psi^\dagger \Gamma_{3001} \Psi \right)^2 + \left( \Psi^\dagger \Gamma_{3032} \Psi \right)^2 \right] \nonumber
\end{equation}
that favors a nematic phase transforming under the $E_g$ representation of the $D_{3d}$ group with $\langle \Psi^\dagger \Gamma_{3001} \Psi \rangle \neq 0$ or $\langle \Psi^\dagger \Gamma_{3032} \Psi \rangle \neq 0$. This phase stands as a candidate for Sym$_{12}$ residing next to PIP$_2$. The sixteen-component Nambu-doubled spinors $\Psi^\dagger$ and $\Psi$ involve spin, valley and sublattice or layer ($b_1$ and $a_2$) degrees of freedom~\cite{supplementary}.

Next we integrate out fast Fourier modes within the Wilsonian shell $\Lambda e^{-\ell} < |\vec{k}|< \Lambda$. Here $\Lambda$ is the ultraviolet momentum cutoff up to which the quasiparticle dispersion remains quadratic and $\ell$ is the logarithm of the RG scale. The coupled RG flow equations are
\allowdisplaybreaks[4]
\begin{eqnarray}~\label{eq:beta_interaction}
\frac{d \lambda_{E_g}}{d \ell}= \lambda^2_{E_g} H(t, \mu, u, h) 
\:\:\: \text{and} \:\:\:
\frac{d x}{d \ell}= 2 x,
\end{eqnarray}    
for $x=t,\mu, u, h$. The relevant Feynman diagrams and $H$ function are shown in the Supplementary Materials (SM)~\cite{supplementary}. The dimensionless quantities are $\lambda_{E_g}=2 m_\star g_{_{E_g}}/(2 \pi)$ and $\tilde{x}=2 m_\star x/\Lambda^2$. For brevity we take $\tilde{x} \to x$. Here $t$ ($h$) is the temperature (Zeeman coupling). The flow equation of $x$ gives an infrared cutoff $\ell^\star_x= \ln [x^{-1}(0)]/2$, where $x(0)<1$ corresponds to its bare value. The RG flow of $\lambda_{E_g}$ terminates at $\ell^\star={\rm min} (\ell^\star_t, \ell^\star_\mu, \ell^\star_u, \ell^\star_h)$. Then the system describes an ordered (a disordered) phase when $\lambda_{E_g}(\ell^\star) >1 \; (<1)$.

Besides the $E_g$ nematicity, singlet $s$-wave and triplet $f$-wave pairings (respectively transforming under the irreducible $A_{1g}$ and $A_{1u}$ representations of the $D_{3d}$ group), nucleate in a degenerate fashion in the ordered state at low temperatures, when $B_\parallel=0$. The $f$-wave pairing changes sign six times among the intralayer next-nearest-neighbor pair of sites [Fig.~\ref{fig:superconductivity}(b)]~\cite{honerkamp:fwave}. But, finite $B_\parallel$ breaks this degeneracy and favors only the spin easy-plane components of the $f$-wave pairing ($\Delta_{A^\perp_{1u}}$). To capture this competition, we allow the conjugate fields, coupling with the corresponding fermion bilinears as 
\allowdisplaybreaks[4]
\begin{eqnarray}
&& \Delta_{E_g} \left[ \Psi^\dagger \Gamma_{3001} \Psi + \Psi^\dagger \Gamma_{3032} \Psi \right], \:\:
\Delta_{A_{1g}} \sum_{\mu=1,2} \Psi^\dagger \Gamma_{\mu 000} \Psi \nonumber \\
&& \Delta_{A^\parallel_{1u}} \sum_{\mu=1,2} \Psi^\dagger \Gamma_{\mu 330} \Psi,  \:\:
\text{and} \:\:
\Delta_{A^\perp_{1u}} \sum_{\mu=1,2} \sum_{j=1,2} \Psi^\dagger \Gamma_{\mu j30} \Psi,
\nonumber 
\end{eqnarray}      
to flow under RG, captured by
\begin{equation}
\frac{d \ln \Delta_{y}}{d \ell} -2=  \lambda_{E_g} J_y (t, \mu, u, h),
\end{equation} 
for $y=E_g$, $A_{1g}$, $A^\parallel_{1u}$, $A^\perp_{1u}$. The relevant Feynman diagrams and $J$ functions are shown in the SM~\cite{supplementary}. Here $\mu=1,2$ reflects U(1) gauge redundancy of the superconducting phase. As $\lambda_{E_g}$ diverges, indicating onset of an ordered phase, the pattern of symmetry breaking is set by the conjugate field that diverges toward $+\infty$ fastest. Following this procedure we construct a few cuts of the phase diagram in the $(\mu, t)$ plane for various $u$ with fixed bare values of $\lambda_{E_g}$ and $h$ [Fig.~\ref{fig:superconductivity}(a)]. The low (high) temperature ordered phase is occupied by easy-plane $f$-wave pairing ($E_g$ nematic metal), manifesting an ``Organizing principle" based on a generalized energy-entropy argument~\cite{szabomoessnerroy:selection, szaboroy:selection}. Namely, the $f$-wave paring that fully gaps the underlying Fermi surface, thereby yielding maximal condensation energy gain, is realized at low temperature. The gapless nematic phase is entropically favored at higher temperature. The matrices involving these two phases fully anticommute with each other and form a composite O(4) vector, exemplifying the ``Selection rule" among competing orders~\cite{royjuricic:selection, szabomoessnerroy:selection, szaboroy:selection}. Such pairing can also be phonon mediated~\cite{dassarma:BLGphonon}.

To anchor the nature of the pairing by approaching from the compensated half-metal (PIP$_2$) side, we compute and compare the bare mean-field susceptibilities ($\chi$) for all symmetry allowed local pairings. By virtue of possessing majority and minority carriers with opposite spin projections, it can sustain both singlet and triplet pairings. For zero external momentum and frequency 
\allowdisplaybreaks[4]	
\begin{eqnarray}~\label{eq:susceptibility}
\hspace{-0.5cm}\chi= -t \sum^{\infty}_{n=-\infty} \int \frac{d^2 \vec{k}}{(2\pi)^2} {\rm Tr} \left[ G(i \omega_n, \vec{k}) M G(i \omega_n, \vec{k}) M \right]. 
\end{eqnarray}
The Hermitian matrix $M$ represents a paired state, fermionic Green's function $G(i \omega_n, \vec{k})=[i \omega_n -H_{\rm CHM}]^{-1}$, $\omega_n=(2 n+1) \pi t$ are the fermionic Matsubara frequencies, and the Boltzmann constant $k_{_B}=1$~\cite{supplementary}. The results are displayed in Fig.~\ref{fig:superconductivity}(c). It shows that $f$-wave pairing always possesses the largest susceptibility and is thus most likely to be nucleated from a parent compensated half-metal. The spin orientation of the paired state gets further locked, such that it is orthogonal to the spin components of both LAF and $B_\parallel$-field.

\emph{Discussions}.~We develop a pedagogical theory for spontaneous symmetry breaking and the resulting degeneracy lifting in BBLG, leading to the formations of quarter- and (compensated) half-metals, as observed in recent experiments~\cite{BLGExpRec:1, BLGExpRec:2}. Our proposals for such fractional metals are equally germane to recently observed quarter- and half-metals in BBLG~\cite{BLGExpRec:3} and RTLG~\cite{young:RTLG1}. The paired state in BBLG, stabilized by applying an in-plane magnetic field, is unambiguously identified to be the easy-plane spin-triplet $f$-wave superconductor. It can emerge from either a nearby parent nematic liquid, mediated by incipient quantum fluctuations, triggered by repulsive electronic interactions or a compensated half-metal.

\emph{Acknowledgments}.~B.R. was supported by a Startup grant from Lehigh University. 



\begin{thebibliography}{}

\bibitem{graphene:RMP} A. H. Castro Neto, F. Guinea, N. M. R. Peres, K. S. Novoselov, and A. K. Geim, Rev. Mod. Phys. {\bf 81}, 109 (2009).

\bibitem{MLG:band1} K. S. Novoselov, A. K. Geim, S. V. Morozov, D. Jiang, M. I. Katsnelson, I. V. Grigorieva, S. V. Dubonos, and A. A. Firsov, Nature (London) {\bf 438}, 197 (2005).

\bibitem{MLG:band2} Y. Zhang, Y. Tan, H. L. Stormer, and P. Kim, Nature (London) {\bf 438}, 201 (2005).

\bibitem{BLG:band} K. S. Novoselov, E. McCann, S. V. Morozov, V. I. Falko, M. I. Katsnelson, U. Zeitler, D. Jiang, F. Schedin, and A. K. Geim, Nat. Phys. {\bf 2}, 177 (2006).

\bibitem{RTLG:band} L. Zhang, Y. Zhang, J. Camacho, M. Khodas, and I. Zaliznyak, Nat. Phys. {\bf 7}, 953 (2011).


\bibitem{TBLGband:1} J. M. B. L. dos Santos, N. M. R. Peres, and A. H. Castro Neto, Phys. Rev. Lett. {\bf 99}, 256802 (2007)

\bibitem{TBLGband:2} R. Bistritzer and A. H. MacDonald, Proc. Natl. Acad. Sci. USA {\bf 108}, 12233 (2011).

\bibitem{TBLGband:3} H.-C. Po, L. Zou, A. Vishwanath, and T. Senthil, Phys. Rev. X {\bf 8}, 031089 (2018). 

\bibitem{TBLGband:4} B. A. Bernevig, Z-D. Song, N. Regnault, and B. Lian, Phys. Rev. B {\bf 103}, 205411 (2021). 


\bibitem{TBLGSC:1} Y. Cao, V. Fatemi, S. Fang, K. Watanabe, T. Taniguchi, E. Kaxiras, and P. Jarillo-Herrero, Nature (London) {\bf 556}, 43 (2018).  

\bibitem{TBLGSC:2} X. Lu, P. Stepanov, W. Yang, M. Xie, M. A. Aamir, I. Das, C. Urgell, K. Watanabe, T. Taniguchi, G. Zhang, A. Bachtold, A. H. MacDonald, and D. K. Efetov, Nature {\bf 574}, 653 (2019).

\bibitem{TBLGSC:3} M. Yankowitz, S. Chen, H. Polshyn, K. Watanabe, T. Taniguchi, D. Graf, A. F. Young, and C. R. Dean, Science {\bf 363}, 1059 (2019).

\bibitem{TBLGSC:4} U. Zondiner, A. Rozen, D. Rodan-Legrain, Y. Cao, R. Queiroz, T. Taniguchi, K. Watanabe, Y. Oreg, F. von Oppen, A. Stern, E. Berg, P. Jarillo-Herrero, and S. Ilani,  Nature (London) {\bf 582}, 203 (2020). 


\bibitem{young:RTLG1} H. Zhou, T. Xie, A. Ghazaryan, T. Holder, J. R. Ehrets, E. M. Spanton, T. Taniguchi, K. Watanabe, E. Berg, M. Serbyn, and A. F. Young, Nature {\bf 598}, 429 (2021).

\bibitem{young:RTLG2} H. Zhou, T. Xie, T. Taniguchi, K. Watanabe, and A. F. Young, Nature {\bf 598}, 434 (2021).


\bibitem{BLGExpRec:1} H. Zhou, Y. Saito, L. Cohen, W. Huynh, C. L. Patterson, F. Yang, T. Taniguchi, K. Watanabe, and A. F. Young, Science {\bf 375}, 774 (2022).

\bibitem{BLGExpRec:2} S. C. de la Barrera, S. Aronson, Z. Zheng, K. Watanabe, T. Taniguchi, Q. Ma, P. Jarillo-Herrero, and R. Ashoori, arXiv:2110.13907


\bibitem{RTLGrecent:0} Y.-Z. Chou, F. Wu, J. D. Sau, and S. Das Sarma, Phys. Rev. Lett. {\bf 127}, 187001 (2021)

\bibitem{RTLGrecent:1} S. Chatterjee, T. Wang, E. Berg, and M. P. Zaletel, arXiv:2109.00002

\bibitem{RTLGrecent:2} A. Ghazaryan, T. Holder, M. Serbyn, and E. Berg, Phys. Rev. Lett. {\bf 127}, 247001 (2021).

\bibitem{RTLGrecent:3} Z. Dong and L. Levitov, arXiv:2109.01133

\bibitem{RTLGrecent:4} T. Cea, P. A. Pantale\'on, V. T. Phong, and F. Guinea, Phys. Rev. B {\bf 105}, 075432 (2022).

\bibitem{RTLGrecent:5} A. L. Szab$\acute{\mbox{o}}$ and B. Roy, Phys. Rev. B {\bf 105}, L081407 (2022).

\bibitem{RTLGrecent:6} Y-Z. You and A. Vishwanath, Phys. Rev. B {\bf 105}, 134524 (2022)


\bibitem{nandkishore:BLG} R. Nandkishore and L. Levitov, Phys. Rev. B {\bf 82}, 115124 (2010).

\bibitem{lemonik:BLG} Y. Lemonik, I. Aleiner, and V. I. Fal'ko, Phys. Rev. B {\bf 85}, 245451 (2012).

\bibitem{macdonald:BLG} F. Zhang, H. Min, and A. H. MacDonald, Phys. Rev. B {\bf 86}, 155128 (2012).

\bibitem{vafek:BLG} V. Cvetkovic, R. E. Throckmorton, and O. Vafek, Phys. Rev. B {\bf 86}, 075467 (2012).

\bibitem{roy:BLG} B. Roy, Phys. Rev. B {\bf 88}, 075415 (2013).


\bibitem{BBLG:nonabelian} R. M. A. Dantas, F. Pe\~na-Benitez, B. Roy, and P. Sur\'owka, Phys. Rev. Research {\bf 2}, 013007 (2020).


\bibitem{supplementary} See Supplementary Materials at XXX-XXXX for explicit Nambu spinor, details of RG and susceptibility analyses, and the effect of trigonal warping. 


\bibitem{katsnelson:hubbard} T. O. Wehling, E. \ifmmode \mbox{\c{S}}\else \c{S}\fi{}a\ifmmode \mbox{\c{s}}\else \c{s}\fi{}\ifmmode \imath \else \i \fi{}o\ifmmode \breve{g}\else \u{g}\fi{}lu, C. Friedrich, A. I. Lichtenstein, M. I. Katsnelson, and S. Bl\"ugel, Phys. Rev. Lett. {\bf 106}, 236805 (2011).


\bibitem{szaboroy:selection} A. L. Szab$\acute{\mbox{o}}$ and B. Roy, Phys. Rev. B {\bf 103}, 205135 (2021). 


\bibitem{interactionstrength:comment} Although in BBLG the strengths of $U$ and $V_2$ are currently unknown, from first principle calculations in monolayer graphene, we expect $U,V_2 \sim$ a few eV, with $U>V_2$~\cite{katsnelson:hubbard}. Here $V_2 \sim e^2/a$ captures a finite range component of Coulomb repulsion, which gets screened due to the metallic gates, where $a$ is distance between $b_1$ or $a_2$ sites. The scale dependent or running or renormalized nematic coupling $g_{_{E_g}}(\ell)=a(\ell)U-b(\ell)V_2$~\cite{szaboroy:selection}, with $a(0)=b(0)=0$, and $\ell$ is the logarithm of the RG time, but $a(\ell) \neq 0$ and $b(\ell) \neq 0$ for any $\ell >0$.


\bibitem{TBLGband:inplanefield} B. Roy and K. Yang, Phys. Rev. B {\bf 88}, 241107(R) (2013).


\bibitem{honerkamp:fwave} C. Honerkamp, Phys. Rev. Lett. {\bf 100}, 146404 (2008).


\bibitem{szabomoessnerroy:selection} A. L. Szab$\acute{\mbox{o}}$, R. Moessner, and B. Roy, Phys. Rev. B {\bf 103}, 165139 (2021). 

\bibitem{royjuricic:selection} B. Roy and V. Juri\v ci\' c, Phys. Rev. B {\bf 99}, 121407(R) (2019). 


\bibitem{dassarma:BLGphonon} Y-Z. Chou, F. Wu, J. D. Sau, and S. Das Sarma, Phys. Rev. B {\bf 105}, L100503 (2022).


\bibitem{BLGExpRec:3} A. M. Seiler, F. R. Geisenhof, F. Winterer, K. Watanabe, T. Taniguchi, T. Xu, F. Zhang, and R. T. Weitz, arXiv:2111.06413

\end{thebibliography}
\end{document}